\documentstyle[preprint,prl,aps]{revtex}

\begin{document}

\bibliographystyle{prsty}

\tighten

\title{Quantum phase gate for photonic qubits\\ using
only beam splitters and post-selection}

\author{Holger F. Hofmann and Shigeki Takeuchi}
\address{CREST, Japan Science and Technology Corporation (JST),\\ 
Research Institute for Electronic Science, Hokkaido 
University,\\
Kita-12 Nishi-6, Kita-ku, Sapporo 060-0812, Japan}

\date{\today}

\maketitle

\begin{abstract}
We show that a beam splitter of reflectivity one-third can be
used to realize a quantum phase gate operation 
if only the outputs conserving the number of photons on
each side are post-selected. 
\end{abstract}

\vspace{0.5cm}

The main difficulty in realizing quantum computation and
quantum information processing for single photon qubits
is the optical implementation of controlled 
interactions between individual qubits. A deterministic
interaction between separate photons would require the
ability to apply strong non-linearities during well-defined
time intervals \cite{Tur95}. At present, there are still many 
technological difficulties preventing the realization of
such reliable quantum gates for photonic qubits.
Recently, however, it has been shown that optical
non-linearities are not necessary for photonic qubit operation
if reliable single photon sources and single photon detectors
are available \cite{Kni01}. The desired interaction between
photonic qubits can also be realized by post-selection. In
particular, quantum phase gates and quantum control-not gates
have been proposed, using additional input ports for single
photons and additional output ports for post-selection
\cite{Kni01,Ral01}. In particular, the post-selection condition
for a successful gate operation requires that the 
number of photons detected in the post-selection ports
is equal to the number previously added to the system.
It is then possible to predict the success of the operation
without further measurements on the output ports. 
In principle, quantum feedback and teleportation could then
be used to realize a scalable optical quantum computer. 
However, the technological difficulties associated with
the requirements of single photon sources and fast feedback
are such that a realization of networks using these
quantum gates is unlikely in the near future.
In this letter, 
we show that a quantum phase gate operation 
can also be achieved using only beam splitters 
if post-selection in the output port is permitted. 
Additional single photon inputs are not necessary, since 
the effective non-linearity can be realized by the
direct interaction of the photonic qubits at the
beam splitter. 
While the operation of 
such a random gate cannot be confirmed without measuring
the output, the fact that single photon sources are
not necessary and that higher efficiency can be achieved with
fewer optical elements should make this proposal an attractive
alternative in the experimental realization of networks
processing multiple photonic qubits. 

The starting point for all linear optics manipulations of
single photon qubits is the unitary operation $\hat{U}_R$
of a beam splitter of reflectivity $R$ on the two mode 
input. In the following, we consider only the four dimensional
Hilbert space associated with zero or one photon in each 
input mode. The unitary operation $\hat{U}_R$ is then
characterized by
\begin{eqnarray}
\hat{U}_R \mid \! 0;0 \rangle &=& \mid \! 0;0 \rangle 
\nonumber \\
\hat{U}_R \mid \! 0;1 \rangle &=& 
\sqrt{R}\mid \! 0;1 \rangle - i 
\sqrt{1-R} \mid \! 1;0 \rangle
\nonumber \\
\hat{U}_R \mid \! 1;0 \rangle &=& 
\sqrt{R}\mid \! 1;0 \rangle - i
\sqrt{1-R} \mid \! 0;1 \rangle
\nonumber \\
\hat{U}_R \mid \! 1;1 \rangle &=& 
(2 R-1) \mid \! 1;1 \rangle - i \sqrt{2R(1-R)}
\left(\mid \! 2;0 \rangle + \mid \! 0;2 \rangle\right).
\end{eqnarray}
Note that the assignment of phases is somewhat arbitrary. 
For convenience, we have chosen the phases so that the 
reflected fields have the same phase as the input fields. 
It is now possible to distinguish photon number conserving
components of the output from components with a changed
photon number distribution in the output. The components
describing photon number conservation are given by the
diagonal elements of the unitary transformation,
\begin{eqnarray}
\langle 0;0 \! \mid \hat{U}_R \mid \! 0;0 \rangle
&=& 1
\nonumber \\ 
\langle 0;1 \! \mid \hat{U}_R \mid \! 0;1 \rangle
&=& \sqrt{R}
\nonumber \\ 
\langle 1;0 \! \mid \hat{U}_R \mid \! 1;0 \rangle
&=& \sqrt{R}
\nonumber \\ 
\langle 1;1 \! \mid \hat{U}_R \mid \! 1;1 \rangle
&=& 2R-1.
\end{eqnarray}
The two photon term is an expression of the destructive
interference between two photon reflection and photon
exchange by two photon transmission. This interference 
originates from the bosonic nature of photons and
causes the preference of two photons in one mode over
two photons in separate modes known as photon bunching.
In order to realize a quantum phase gate, this interference
may be applied to create a phase change in the two photon
case. Specifically, the two photon term of beam splitters 
with $R<1/2$ is dominated by the negative amplitude of
mutual photon transmission, causing a phase change of $\pi$
with respect to the case of mutual
reflection dominating at $R>1/2$. Since photon number 
conservation requires reflection for both single photon
inputs, post-selecting only those cases where the output 
photon number equals the input photon number ensures that
the phase shift associated with photon transmission only
occurs for the two photon input. Post-selection thus
converts beam splitters with $R<1/2$ into natural phase
gate elements. The only problem remaining is that the 
post selection also changes the amplitudes associated
with each input component, indicating different probabilities
of success for different inputs. 

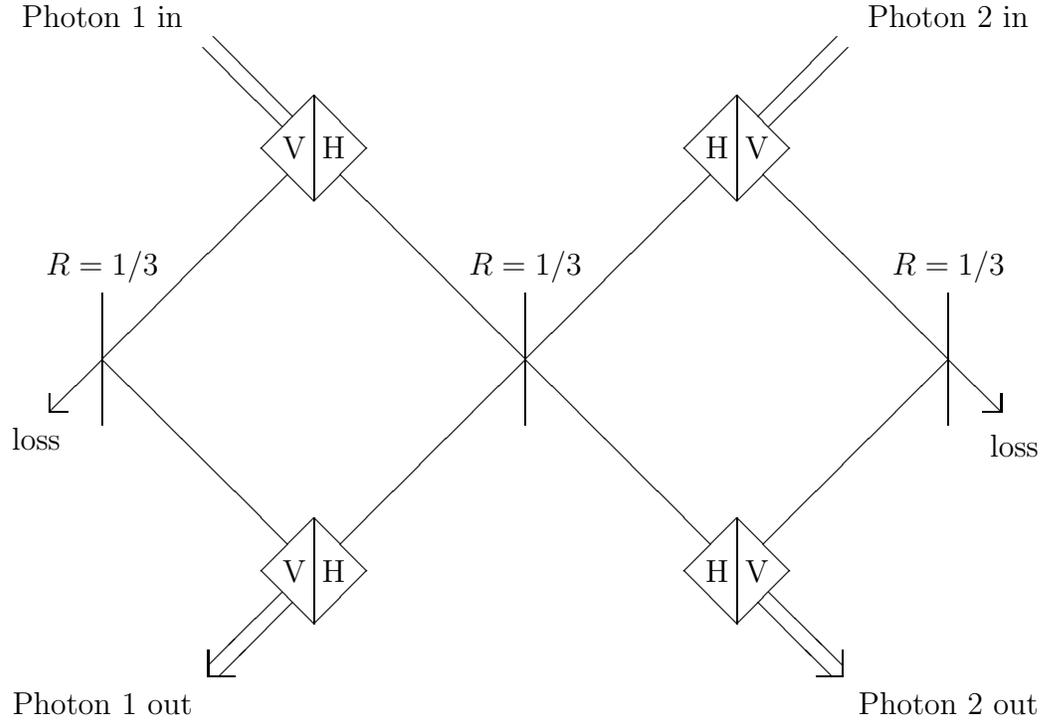
\begin{figure}
\begin{picture}(400,320)

\put(108,248){\line(-1,1){30}}
\put(112,252){\line(-1,1){30}}
\put(0,280){\makebox(80,20){Photon 1 in}}
\put(120,220){\line(0,1){40}}
\put(120,260){\line(1,-1){20}}
\put(120,260){\line(-1,-1){20}}
\put(120,220){\line(1,1){20}}
\put(120,220){\line(-1,1){20}}
\put(105,230){\makebox(15,20){V}}
\put(120,230){\makebox(15,20){H}}
\put(110,230){\line(-1,-1){90}}
\put(20,140){\line(0,1){7}}
\put(20,140){\line(1,0){7}}
\put(0,120){\makebox(30,20){loss}}
\put(110,90){\line(-1,1){70}}
\put(20,185){\makebox(40,20){$R=1/3$}}
\put(40,135){\line(0,1){50}}

\put(120,60){\line(0,1){40}}
\put(120,100){\line(1,-1){20}}
\put(120,100){\line(-1,-1){20}}
\put(120,60){\line(1,1){20}}
\put(120,60){\line(-1,1){20}}
\put(105,70){\makebox(15,20){V}}
\put(120,70){\makebox(15,20){H}}
\put(108,72){\line(-1,-1){28}}
\put(112,68){\line(-1,-1){28}}
\put(80,40){\line(1,0){10}}
\put(80,40){\line(0,1){10}}
\put(0,20){\makebox(80,20){Photon 1 out}}


\put(292,248){\line(1,1){30}}
\put(288,252){\line(1,1){30}}
\put(320,280){\makebox(80,20){Photon 2 in}}
\put(280,220){\line(0,1){40}}
\put(280,260){\line(-1,-1){20}}
\put(280,260){\line(1,-1){20}}
\put(280,220){\line(-1,1){20}}
\put(280,220){\line(1,1){20}}
\put(265,230){\makebox(15,20){H}}
\put(280,230){\makebox(15,20){V}}
\put(290,230){\line(1,-1){90}}
\put(380,140){\line(0,1){7}}
\put(380,140){\line(-1,0){7}}
\put(370,120){\makebox(30,15){loss}}
\put(290,90){\line(1,1){70}}
\put(340,185){\makebox(40,20){$R=1/3$}}
\put(360,135){\line(0,1){50}}

\put(280,60){\line(0,1){40}}
\put(280,100){\line(-1,-1){20}}
\put(280,100){\line(1,-1){20}}
\put(280,60){\line(-1,1){20}}
\put(280,60){\line(1,1){20}}
\put(265,70){\makebox(15,20){H}}
\put(280,70){\makebox(15,20){V}}
\put(292,72){\line(1,-1){28}}
\put(288,68){\line(1,-1){28}}
\put(320,40){\line(-1,0){10}}
\put(320,40){\line(0,1){10}}
\put(320,20){\makebox(80,20){Photon 2 out}}

\put(270,90){\line(-1,1){140}}
\put(130,90){\line(1,1){140}}
\put(180,185){\makebox(40,20){$R=1/3$}}
\put(200,135){\line(0,1){50}}

\end{picture}

\caption{\label{setup} Schematic setup of the post-selected
quantum phase gate. Vertical lines represent beam splitters
of reflectivity 1/3, boxes represent polarization beam 
splitters transmitting horizontally polarized light and
reflecting vertically polarized light.}
\end{figure}

In order to compensate the imbalance between the zero, one,
and two photon inputs, the reduction in amplitude should
correspond to linear photon losses. In other words, the two
photon amplitude of $|2R-1|$ should
be equal to the square of the one photon amplitudes 
$|\sqrt{R}|$. This condition is fulfilled for beam splitters
with one third reflectivity. 
The post-selected transformation at the beam splitter now
reads
\begin{eqnarray}
\label{eq:1/3}
\langle 0;0 \! \mid \hat{U}_{1/3} \mid \! 0;0 \rangle
&=& \hspace{0.5cm} 1
\nonumber \\ 
\langle 0;1 \! \mid \hat{U}_{1/3} \mid \! 0;1 \rangle
&=& \sqrt{1/3}
\nonumber \\ 
\langle 1;0 \! \mid \hat{U}_{1/3} \mid \! 1;0 \rangle
&=& \sqrt{1/3}
\nonumber \\ 
\langle 1;1 \! \mid \hat{U}_{1/3} \mid \! 1;1 \rangle
&=& - 1/3.
\end{eqnarray}
This operation is equal to a non-linear phase shift of
$\pi/2$ for the two photon case and a loss of two thirds
of the input photons. It is then possible to construct a
phase gate for photonic qubits by applying this operation
e.g. to the horizontally (H) polarized components of the 
qubits, while balancing the losses by also attenuating
the vertically (V) polarized components to $1/3$ of their
input photon number. Figure \ref{setup} shows the complete
post-selected phase gate setup. If there is exactly one
photon of arbitrary polarization in each input port, and
if there is exactly one photon of any arbitrary polarization 
in each output port, this simple setup is described by 
the quantum phase gate operation 
$\hat{S}_{\mbox{QPG}}$ with
\begin{eqnarray}
\hat{S}_{\mbox{QPG}} \mid \! V;V\rangle &=& 
\hspace{0.5cm} \frac{1}{3} \mid \! V;V\rangle
\nonumber \\
\hat{S}_{\mbox{QPG}} \mid \! V;H\rangle &=& 
\hspace{0.5cm} \frac{1}{3} \mid \! V;H\rangle
\nonumber \\
\hat{S}_{\mbox{QPG}} \mid \! H;V\rangle &=& 
\hspace{0.5cm} \frac{1}{3} \mid \! H;V\rangle
\nonumber \\
\hat{S}_{\mbox{QPG}} \mid \! H;H\rangle &=& 
\hspace{0.2cm} -  \frac{1}{3} \mid \! H;H\rangle.
\end{eqnarray}
The application as a quantum controlled-NOT is also 
straightforward. Either the polarization of one input 
port is rotated by $\pi/4$, or the coding is defined
separately in port one and port two, such that
\begin{equation}
\begin{array}{rclcrcl}
\mid\! 0 \rangle_1 &=& \mid\! V \rangle_1 &\hspace{1cm}&
\mid\! 0 \rangle_2 &=& (\mid\! V \rangle_2 + \mid\! H \rangle_2)
\\
\mid\! 1 \rangle_1 &=& \mid\! H \rangle_1 &&
\mid\! 1 \rangle_2 &=& (\mid\! V \rangle_2 - \mid\! H \rangle_2).
\end{array}
\end{equation}
In terms of the quantum bit states defined by this encoding,
the operation of the same setup then reads
\begin{eqnarray}
\hat{S}_{\mbox{QPG}} \mid \! 00 \rangle &=& 
\hspace{0.5cm} \frac{1}{3} \mid \! 00 \rangle
\nonumber \\
\hat{S}_{\mbox{QPG}} \mid \! 01 \rangle &=& 
\hspace{0.5cm} \frac{1}{3} \mid \! 01 \rangle
\nonumber \\
\hat{S}_{\mbox{QPG}} \mid \! 10 \rangle &=& 
\hspace{0.5cm} \frac{1}{3} \mid \! 11 \rangle
\nonumber \\
\hat{S}_{\mbox{QPG}} \mid \! 11 \rangle &=& 
\hspace{0.5cm}  \frac{1}{3} \mid \! 10 \rangle.
\end{eqnarray}
In all cases, the attenuation factor of $1/3$ indicates an
efficiency of $1/9$ for the post-selected gate operation.
Note that this compares somewhat favorably with the efficiency
of $1/16$ obtained in previous proposals \cite{Kni01,Ral01}.
However, the main problem of our simplified quantum gate is
not so much the efficiency itself, but the difficulty of 
determining the post-selection condition. As opposed to the
previous proposals, the setup discussed here requires 
post-selection in the output ports themselves. Some consideration
should therefore be given to the kind of errors, and to the 
possible strategies of error identification.

In principle, two types of errors should be distinguished:
photon bunching errors where both input photons are found
in the same output port, and photon loss errors where one or
both photons do not reach the output ports at all. 
The setup shows that photon loss errors are only possible for 
the V-polarized components. The chances for photon losses are
$8/9$ for the $\mid \! V;V \rangle$ component, $2/3$ for
the $\mid \! V;H \rangle$ and $\mid \! H;V \rangle$ components,
and zero for the $\mid \! H;H \rangle$ component. Photon losses 
can be detected by high efficiency photon detectors 
\cite{Tak99} placed in the respective output ports of the 
beam splitters inducing the 
losses. On the other hand, bunching errors are caused by
the H-polarized components. For the $\mid \! H;H \rangle$
component, the probability of photon bunching at the central
beam splitter is $8/9$. For the $\mid \! V;H \rangle$ and 
$\mid \! H;V \rangle$ components, there is a $2/9$ probability
that the H-polarized photon will be transmitted to the side of
the V-polarized qubit while the V-polarized photon is not lost,
resulting in an effective bunching error in that qubit.
Bunching errors can only be detected by measurements in the 
output ports. However, since only two photons entered the 
setup, and since no additional photons were generated, 
it is sufficient to measure the total photon number in only
one of the two output ports. 
In a network of quantum gates, it is necessary to keep track
of the propagation of bunching errors, since it is possible that
the photons become unbunched again at some point, causing 
undetectable errors in the total network operation. 
Within a larger network, it may therefore be most efficient to 
combine the post-selected gate proposed here with the previous
proposals \cite{Kni01,Ral01} for an optimized total efficiency of
the network. In particular, it may be useful to apply this
quantum gate in the last stages, where post-selection in the
output is not so problematic.
 
At the present stage of development, the post-selected gate 
proposed here may already be sufficient to realize small
multi-qubit networks. Up to now, most operations
on photonic qubits have not gone beyond qubit pairs, and even
the generation of three or four photon entanglement has only
been achieved by post-selection in the outputs \cite{Bou99,Pan01}. 
This is related to the fact that the most common method of
generating photon number states is by unpredictable
spontaneous downconversion events. When this method of photon
generation is used, the rate of coincidences between
multiple emission events drops exponentially as more and
more events are required. It may therefore be extremely 
difficult to implement networks using optical quantum gates
that require additional photon inputs such as the ones 
proposed in \cite{Kni01,Ral01} with conventional technology. 
The post-selected quantum gate
proposed here minimizes the number of coinciding photons
by requiring only the reliable generation of an
input pair carrying the quantum information. An N-qubit
network could then be realized using an N-pair coincidence
in the downconversion processes. 

In conclusion, the post-selected quantum gate is more simple to 
implement and more efficient than previous proposals.
It could thus contribute to the realization of multi-qubit 
processing by significantly reducing the technological 
requirements for optical quantum gate operations.

\end{document}